\documentclass[conference]{IEEEtran}
\IEEEoverridecommandlockouts

\usepackage{graphicx} 
\usepackage{mathtools}
\usepackage{amsthm}
\usepackage{amssymb}
\usepackage{tikz}
\usepackage{tikz-cd}
\usepackage{braket} 
\usepackage{hyperref}
\usepackage[capitalize]{cleveref}
\usepackage{multirow}
\newtheorem{theorem}{Theorem}

\newtheorem{definition}{Definition}
\newtheorem{corollary}{Corollary}
\newtheorem{proposition}{Proposition}
\newtheorem{remark}{Remark}
\newtheorem{example}{Example}

\newcommand\ie{\textit{i.e.,~}}
\newcommand\eg{\textit{e.g.,~}}

\newcommand{\wt}{{\textrm wt}}
\newcommand{\en}{{\textrm En}}
\newcommand{\Span}{{\textrm Span }}
\newcommand{\Supp}{{\textrm Supp}}
\newcommand{\F}{\mathbb{F}}

\newcommand{\defeq}{\vcentcolon=}

\title{Asymptotically good CSS-T codes and a new construction of triorthogonal codes\\
\thanks{This work was supported by the Spanish Ministry of Economy and Competitiveness through the MADDIE project (Grant No. PID2022-137099NBC44), by the Spanish Ministry of Science and Innovation through the project “Few-qubit quantum hardware, algorithms and codes, on photonic and solidstate systems" (PLEC2021-008251), by the Ministry for Digital Transformation and of Civil Service of the Spanish Government through the QUANTUM ENIA project call - QUANTUM SPAIN project, and by the European Union through the Recovery, Transformation and Resilience Plan - NextGenerationEU within the framework of the Digital Spain 2026 Agenda, and by the grants ANR-21-CE39-0009-BARRACUDA and ANR-22-CPJ2-0047-01 from the French National Research Agency. J.E.M. is funded by the Spanish Ministry of Science, Innovation and Universities through a Jose Castillejo mobility grant for his stay at the Cavendish Laboratory of the University of Cambridge.
}
}

\author{\IEEEauthorblockN{Elena Berardini}
\IEEEauthorblockA{\textit{Mathematical Institute of Bordeaux} \\
\textit{CNRS, University of Bordeaux}\\
Talence, France \\
elena.berardini@math.u-bordeaux.fr}
\and
\IEEEauthorblockN{Reza Dastbasteh}
\IEEEauthorblockA{\textit{Department of Basic Sciences} \\
\textit{Tecnun - University of Navarra}\\
San Sebastian, Spain \\
rdastbasteh@unav.es}
\and
\IEEEauthorblockN{Josu Etxezarreta Martinez}
\IEEEauthorblockA{\textit{Department of Basic Sciences} \\
\textit{Tecnun - University of Navarra}\\
San Sebastian, Spain \\
jetxezarreta@unav.es}
\and
\IEEEauthorblockN{Shreyas Jain}
\IEEEauthorblockA{\textit{Department of Mathematical Sciences} \\
\textit{Indian Institute of Science Education and Research}\\
 Mohali, India \\
 \textit{Mathematical Institute of  Bordeaux} \\
\textit{CNRS, University of Bordeaux}\\
Talence, France \\
ms20098@iisermohali.ac.in}
\and
\IEEEauthorblockN{Olatz Sanz Larrarte}
\IEEEauthorblockA{\textit{Department of Basic Sciences} \\
\textit{Tecnun - University of Navarra}\\
San Sebastian, Spain  \\
osanzl@unav.es}
}

\def\BibTeX{{\rm B\kern-.05em{\sc i\kern-.025em b}\kern-.08em
    T\kern-.1667em\lower.7ex\hbox{E}\kern-.125emX}}

\begin{document}
\sloppy
\maketitle
\begin{abstract}
    We propose a new systematic construction of CSS-T codes from any given CSS code using a map $\phi$. When $\phi$ is the identity map $I$, we retrieve the construction of \cite{hu2021mitigating} and use it to prove the existence of asymptotically good binary CSS-T codes, resolving a previously open problem in the literature, and of asymptotically good quantum LDPC CSS-T codes. 
    We analyze the structure of the logical operators corresponding to certain non-Clifford gates supported by the quantum codes obtained from this construction ($\phi = I$), concluding that they always result in the logical identity. An immediate application of these codes in dealing with coherent noise is discussed. 
    We then develop a new doubling transformation for obtaining triorthogonal codes, which generalizes the doubling construction presented in \cite{jain2024}. Our approach permits using self-orthogonal codes, instead of only doubly-even codes, as building blocks for triorthogonal codes. This broadens the range of codes available for magic state distillation.
\end{abstract}


\section{Introduction}

Quantum error correction is a well-known and widely used technique for protecting quantum information from corruption by noise and, consequently, enabling the possibility of constructing fault-tolerant quantum computers. An $[\![n,k,d]\!]$ binary quantum error-correcting code encodes $k$ logical qubits of information into $n$ physical qubits and can correct up to $\lfloor \frac{d-1}{2} \rfloor$ errors. By itself, a quantum error correction code acts as a memory. However, operations over the logical level are required to obtain fault-tolerant quantum computers capable of realizing tasks that may exhibit quantum advantage. In this sense, logical operations are realized by manipulating the physical qubits encoding the information so that the desired logical operator is applied to the encoded logical qubits. Importantly, the operations applied to the physical qubits should preserve the code space; otherwise, how the logical qubit is protected would essentially be lost. Furthermore, those logical operators must be induced in a fault-tolerant manner, implying that the overall quantum computation on the logical level is reliable enough.

The most natural approach to fault tolerance is the design of transversal gates, which are formed by the tensor product of single-qubit gates.
A key feature of transversal operators is that single errors do not propagate within the code blocks during their operation, making errors trackable.

In general, the Eastin--Knill theorem states that no universal gate set can be implemented on the logical level in a transversal manner for binary quantum error-correcting codes with $d>1$ \cite{eastin2009restrictions}. 
Recently, a way to circumvent such a no-go theorem has been proposed under certain restrictions on the error model, but its application seems to be restricted to $d$-dimensional quantum systems \cite{EastinCircumvent}. 
There exist several universal quantum gate sets, with the Clifford+$T$ gate set being the most studied one \cite{solovayKitaev}. 
Quantum circuits composed purely of Clifford gates are efficiently classically simulable as a result of the Gottesman--Knill theorem \cite{Gottesman:1998hu}. Implementing only transversal Clifford gates is both easier and less resource-intensive. In fact, finding codes accepting logical Clifford gates in a transversal manner is relatively straightforward. 
For example, any Calderbank--Shor--Steane (CSS) code constructed from a doubly-even code, and its dual, admits transversal Clifford operators \cite{dastbasteh2024quantum,grassl2000cyclic,S99_nature}.
Nevertheless, those codes would lack non-Clifford gates, implying that the quantum computations that can be realized would be classically simulable. A way to circumvent this problem is by means of magic state distillation and injection \cite{Campbell2017}. 
Magic state distillation protocols are closely connected to quantum error correction codes that admit the transversal implementation of a non-Clifford gate, \ie such an operation does not change the code space. Note that for such a code to be useful for magic state distillation, the transversal implementation should not only preserve the codespace but also result in a non-Clifford operation at the logical level. A common family of such codes is that of triorthogonal codes \cite{bravyi2012magic}.

Motivated by this application, numerous works in the literature have focused on constructing and characterizing stabilizer codes that support transversal $T$ gates; see, for instance, \cite{andrade2023css,BCR24,CMLMRS24,camps2024binary,RCNP20,RCNP20b}. 
In particular, \cite{RCNP20} introduces CSS codes supporting physical transversal $T$ gates, known as CSS-T codes. 
The applicability of CSS-T codes to magic state distillation depends on the induced logical operators. 
Furthermore, quantum codes supporting physical transversal $T$ and finer rotations over the $Z$ axis that result in a trivial action on the logical space, \ie logical identity, are useful to deal with coherent errors at the physical level \cite{hu2021mitigating}. A limitation of all currently known infinite families of CSS-T codes is that they exhibit either a vanishing rate or a vanishing relative distance \cite{andrade2023css, jain2024,scruby2024}.
An intriguing open problem, proposed in \cite{RCNP20}, is to construct asymptotically good CSS-T codes, which simultaneously achieve a non-vanishing rate and relative distance.

\paragraph{Our contribution.} Our main contribution is twofold. First, motivated by the open problem mentioned above, we introduce a construction that produces a binary CSS-T code from any given binary CSS code using a map $\phi$ respecting some properties. When $\phi$ is the identity map, we retrieve the construction of \cite{hu2021mitigating} and employ it to prove the existence of asymptotically good binary CSS-T codes. 
Additionally, we show that this construction preserves the low-density parity check (LDPC) property, proving the existence of asymptotically good LDPC codes that are also CSS-T. We also demonstrate that our method's natural generalization yields asymptotically good CSS codes that support transversal gates corresponding to any $Z$ rotation angle.
Furthermore, we study the structure of the logical operators corresponding to the transversal $T$ gate, finer $Z$ rotations, and the $CCZ$ gate on our proposed CSS codes, which have the desired symmetry for these transversal gates. 
In particular, we prove that all such transversal gates act as the logical identity operator when applied to the CSS codes from our construction.
However, these codes have applications in mitigating coherent noise, and we discuss this application.

Secondly, we present a natural method for transforming certain CSS-T codes into triorthogonal codes. This result allows us to generalize the doubling construction discussed in \cite{jain2024} by permitting the use of general self-orthogonal codes, not necessarily self-dual, as input to the construction. This new perspective broadens the range of codes that can be used to generate triorthogonal codes, potentially enhancing the applicability of CSS codes in magic state distillation. In particular, we provide an example of a triorthogonal code that can be generated using our approach but cannot be obtained using the methods discussed in \cite{jain2024}.

\paragraph{Organization of the paper.} In Section \ref{sec:preliminaries}, we set the notations and give definitions and known results on classical linear codes and quantum gates and codes. 
We also recall the definitions of CSS, CSS-T  and triorthogonal codes, which are the main focus of this paper. 
Section \ref{sec:CCSTconstruction} presents our construction of CSS-T codes from CSS ones and their properties. In particular, we prove here the existence of asymptotically good CSS-T codes.
In Section \ref{sec:transversalgates}, we discuss how to generalize our construction to support other transversal gates and examine their corresponding logical operators.
Section \ref{sec:triorthogonal} is devoted to our new construction of triorthogonal codes. Finally, in Section \ref{sec:conclusion}, we wrap up our work and present future research questions.

\section{Preliminaries}\label{sec:preliminaries}
In this section, we give a short introduction to classical and quantum error-correcting codes and their interplay. 
This allows us to fix the notations used for the rest of the paper. 
We refer the reader to \cite{HP10} and to \cite{gottesman1997stabilizer,NC02} for a detailed presentation of classical and quantum codes, respectively. 
\subsection{Classical linear codes}
We work over the binary field  $\F_2$. A (\emph{linear}) \emph{code}
$C$ is an $\F_2$-linear subspace of $\F_2^n$, where $n$ is a positive integer. 
A vector $x\in C$ will be called a \emph{codeword}. The \emph{length} of the code $C$ is the dimension of the ambient space, namely $n$. The \emph{dimension} of $C$ is its dimension as a linear subspace over $\F_2$. 
A linear code $C$ can be defined using a {\em generator matrix}, that is, a binary matrix with the row space equal to $C$. 
A linear code $D \subseteq C$ is called a \emph{subcode} of $C$. 

The \emph{support} of a vector $x \in \F_2^n$  is the set $\Supp(x):=\{i \in \{1,\dots, n\} \, | \, x_i \neq 0\}$.
The (\emph{Hamming}) \emph{weight} of a codeword $x \in C$ is defined by 
$\wt(x) := |\Supp(x)|$. If $\wt(x)\equiv 0 \pmod 2$, then $x$ is called an {\em even (weight) vector}. A linear code $C\subseteq \F_2^n$ is called an {\em even code} if it only contains even vectors. For a set of codewords $S$ we use the notation $\wt(S)\coloneqq\{\wt(x) \mid x\in S, x\neq \mathbf{0}\}$, where $\boldsymbol{0}$ denotes the all-zeros vector.
Finally, the \emph{minimum distance} $d(C)$ of a code $C$
is defined as
$d(C)\coloneqq\min\{\wt(x) \mid x \in C, \, x \neq \boldsymbol{0}\}.$
 We will refer to a code of length $n$, dimension $k$ and minimum distance $d$ as an $[n,k,d]$ code.

The \emph{(Euclidean) dual} of $C\subseteq\F_2^n$ is the linear code
$C^\perp\coloneqq\{ y\in\F_2^n \mid x \cdot y=0 \mbox{ for all } x \in C\},$
where $x \cdot y$ denotes the {\em Euclidean inner product} of the vectors $x$ and $y$.
If $C=C^\perp$, the code is called {\em self-dual}.
The component-wise multiplication of $x,y\in\F_2^n$, or the {\em Schur product}, is 
$$(x_1,x_2,\ldots,x_n)\star(y_1,y_2,\ldots,y_n)\coloneqq(x_1y_1,x_2y_2,\ldots,x_n y_n).$$
The Schur product of two codes $C,D\subseteq\F_2^n$ is then defined as
$C\star D\coloneqq \Span\{x\star y \mid x\in C, y\in D\}.$
The Schur product $C\star C$ is denoted $C^{\star 2}$.  
One can easily see that $C\subseteq C^{\star 2}$. 
We refer to \cite{Randriam15} for an extensive survey on the Schur product of codes.
\subsection{Quantum gates and quantum error correction}

Throughout the paper, $\mathcal{H}$ is the complex Hilbert space $\mathbb{C}^2$ endowed with the inner product
$$\langle u,v \rangle=\sum_{i=1}^{n} \overline{u_i} v_i,$$
where $v,u \in \mathcal{H}$ and $\overline{u_i}$ is the complex conjugate. A \emph{qubit} of length one (or $1$-qubit) is an element of $\mathcal{H}$ of norm one. 
We show the set of one-qubit Pauli operators by $I,X,Y,$ and $Z$.
Let $\{\ket 0,\ket 1\}$ be a basis for $\mathcal{H}$, where $\ket 0=\begin{bmatrix} 1 \\ 0 \end{bmatrix}$ and 
$\ket 1=\begin{bmatrix} 0 \\ 1 \end{bmatrix}$. An arbitrary state of a closed 1-qubit quantum system is defined by 
$\ket \varphi=\alpha \ket 0+\beta \ket 1,$ 
where $\alpha,\beta \in \mathbb{C}$ and $|\alpha|^2+|\beta|^2=1$.
The action of each of the Pauli operators $P$ on $\ket{\varphi}$ is defined by $P\ket{\varphi}$. 
Recall that the $Z$ rotation by the angle $\frac{2\pi}{2^{\ell}}$ for some $\ell \ge 0$ is given by the gate
$$\varGamma^{(\ell)}=\begin{bmatrix}1&0\\0&e^{\frac{2\pi i}{2^\ell}}\end{bmatrix}.$$ 
Then, the action of $\varGamma^{(\ell)}$ on the basis states is 
$\varGamma^{(\ell)} |0\rangle \longmapsto |0\rangle$ and $\varGamma^{(\ell)} |1\rangle \longmapsto \mathrm{e}^{\frac{2\pi i}{2^{\ell}}}|1\rangle.$
If we fix $\ell = 0,1,2,3$, then the $Z$ rotation $\varGamma^{(\ell)}$ is in correspondence to $I,Z,S$, and $T$, respectively.  

For any $n\in\mathbb{N}$, we let $\mathcal{H}^n=\mathcal{H}\otimes \dots \otimes \mathcal{H}$ to be the $2^n$-dimensional complex Hilbert space $\mathbb{C}^{2^n}$, where $\otimes$ denotes the usual tensor product of elements in $\mathbb{C}^2$. 
Qubits of length $n$ (or $n$-qubits) are elements of $\mathcal{H}^n$ of norm one. For the rest of this paper, we fix $\{\ket a \mid a\in \F_2^n \}$ to be a basis for $\mathcal{H}^{n}$.

\subsection{From classical to quantum codes}
We start by reviewing the construction of binary quantum CSS codes from a pair of classical codes.
\begin{theorem}[\cite{CS96}]\label{th:css}
Let $C_2,C_1 \subseteq \mathbb{F}_2^n$  be two linear codes such that $C_2 \subseteq C_1$. Then, there exists an $[\![n,k,d]\!]$ quantum code, called CSS and denoted by $(C_1,C_2)_\mathrm{CSS}$, where $k=\dim(C_1)-\dim(C_2)$ and $d=\min\{\wt(C_1\setminus C_2),\wt(C_2^\perp\setminus C_1^\perp)\}$.
\end{theorem}
The quantum CSS code of \cref{th:css} has a {\em parity check matrix} of the form
$$\begin{bmatrix}
        H_X & 0\\
        0 & H_Z
\end{bmatrix},$$
where $H_X$ and $H_Z$ are generator matrices of $C_2$ and $C_1^\perp$, respectively. 

Throughout the paper, we will use the notation $d_X$ and $d_Z$ for $\min\{\wt(C_1\setminus C_2)\}$ and $\min\{\wt(C_2^\perp\setminus C_1^\perp)\}$, respectively, and refer to them as the $X$-distance and the $Z$-distance, respectively. 
Note that since $d_X\geq \min\left\{d(C_1)\right\}$ 
 and $d_Z\geq \min\left\{d(C_2^\bot)\right\}$, we always have $d\geq \min\left\{d(C_1),d(C_2^\bot)\right\}$ for a quantum code $(C_1,C_2)_\mathrm{CSS}$. 
When equality holds in the latter inequality, we say that the quantum CSS code is {\em non-degenerate}. Otherwise, the code will be called {\em degenerate}.

Let $C_2 \subseteq C_1$ be binary linear codes of length $n$ that define a binary quantum CSS code with parameters $[\![n,k,d]\!]$. 
Let $C_1=C_2\oplus \Span\{x_1,x_2,\ldots,x_k\}$, for some $x_1,x_2,\ldots,x_k \in C_1 \setminus C_2$. Let $H$ be a $k \times n$ binary matrix having $x_1,x_2,\ldots,x_k$ as its rows.  
A common encoding approach for this binary quantum code is defined by $\en:\mathcal{H}^{k} \rightarrow \mathcal{H}^{n}$, where for $u\in \F_2^{k}$ we have
\begin{equation}\label{encoding}
\en \ket{u}=\frac{1}{\sqrt{|C_2|}}\sum_{x\in C_2} \ket{x+uH}.
\end{equation} 
Note that the quantum state $\ket u$ on the left-hand side of Equation \eqref{encoding} is a $k$-qubit state, while the right-hand side quantum states are $n$-qubit states. To distinguish them, from now on, we show the logical states using the notation $\ket{u}_L$. Moreover, to simplify our computations, {\em we discard the normalization factor}.  
Each operation on such a quantum code is in the form of an $n$-qubit operation that preserves the code space, \ie $\en(\mathcal{H}^{k})$. 

Recently, Rengaswamy \emph{et al.~}characterized all stabilizer codes in which the transversal $T$ gate, and some other variants of it, preserve the code space \cite{RCNP20b}. 
In \cite{RCNP20b}, it was also shown that CSS codes are optimal among all non-degenerate stabilizer codes with this property. 
A restriction of the characterization from \cite{RCNP20b} to CSS codes, commonly known as {\em CSS-T codes}, is stated below. 
\begin{definition}[\cite{RCNP20}]\label{def:csst}
A CSS code $(C_1,C_2)_\mathrm{CSS}$ is called a \emph{CSS-T code} if the following hold:
\begin{enumerate}
    \item the code $C_2$ is even, \ie$\wt(x) \equiv 0 \pmod 2$ for each $x\in C_2$;
\item for each $x\in C_2$, there exists a self-dual code in $C_1^\perp$ of dimension $\wt(x)/2$ that is supported on $x$, \ie there exists $C_x \subseteq C_1^\perp$ such that $|C_x| = 2^{\wt(x)/2}$, $C_x = C_x^{\perp_x}$, where $\perp_x$ denotes the dual taken in $\F_2^{\wt(x)}$, and for each $z \in C_x$ we have $\Supp(z) \subseteq \Supp(x)$.
\end{enumerate}
\end{definition}
Importantly, a CSS-T code is defined as a CSS code for which a transversal application of $T$ gates at the physical level, \ie $T^{\otimes n}$, preserves the codespace. In this sense, the resulting logical operator for CSS-T codes can be either the logical identity or the logical $T$ after the application of $T^{\otimes n}$ at the physical level.
A reformulation of the above conditions in terms of the Schur product of the ingredient classical codes was proved in \cite{CMLMRS24}, and we recall below.
\begin{theorem}[{\cite[Theorem 2.3]{CMLMRS24}}]\label{th:csstchar}
 A pair of binary linear codes $C_1, C_2$ generates a CSS-T code if and only if $C_1^\perp + C_1^{\star 2} \subseteq C_2^\perp$.
\end{theorem}
If we consider a pair of linear codes $C_2\subseteq C_1$ giving a CSS code, we obtain the inclusion $C_1^\perp \subseteq C_2^\perp$. 
Therefore, each given CSS code $(C_1,C_2)_\mathrm{CSS}$, is a CSS-T code if and only if $C_1^{\star 2} \subseteq C_2^\perp$.
Note also that for each $ x, y \in C_1 $, the support of the Schur product $ x \star y $ is the set of coordinate positions where both $ x $ and $ y $ are nonzero (also known as the intersection of corresponding stabilizers). 
Note that a CSS-T code may have stabilizer generators with negative signs. As mentioned in \cite[Remark 3]{RCNP20}, applying a Pauli correction can transform the code space into the form of Equation \cref{encoding}. Thus, without loss of generality, we assume this encoding.

Some constructions and examples of CSS-T codes are provided in \cite{andrade2023css,CMLMRS24,camps2024binary,RCNP20,RCNP20b}. 
All currently known families of CSS-T codes have either a vanishing rate or a vanishing relative distance. 
An interesting open problem, which will be solved in the next section, is constructing an infinite family of asymptotically good CSS-T codes with non-vanishing rate and relative distance \cite{RCNP20}.
\subsection{Triorthogonal codes}
Quantum triorthogonal codes are a subclass of binary CSS-T codes which can realize the logical  $T$ gate in a transversal manner (up to Pauli corrections). Many magic state distillation protocols rely on such codes, see \eg \cite{bravyi2012magic, litinski2, lee2024low}. 
Here, we briefly recall the definition of binary triorthogonal codes. 
\begin{definition}[Triorthogonal Matrices, \cite{bravyi2012magic}]
A binary matrix $G$ of size $m \times n$ is called triorthogonal if and only if the supports of any pair and any triple of its distinct rows have even overlap, \ie
$$\sum_{j=1}^nG_{a,j}G_{b_j} \equiv 0\mod{2}$$
for all pairs of rows $1 \leq a < b \leq m$ and
$$\sum_{j=1}^nG_{a,j}G_{b_j}G_{c_j} \equiv 0\mod{2}$$
for all triples of rows $1 \leq a < b < c\leq m.$
\end{definition}
We call the linear code generated by a triorthogonal matrix a \emph{triorthogonal subspace}. 
We can write a triorthogonal matrix as $G=[G_1,G_0]^T$,
where $G_1$ consists of odd-weight rows, and $G_0$ consists of even-weight rows. 
\begin{theorem}[Triorthogonal Codes \cite{bravyi2012magic}]
Let $G=[G_1,G_0]^T$ be a triorthogonal matrix, where $G_1$ consists of all $k$ odd-weight rows.
Let $[\![n,k]\!]$ be the CSS code of $(C_1,C_2)$ formed when $G_0$ and $G$ (with $G \neq G_0$) are taken as the generator matrices of $C_2$ and $C_1$, respectively. Then applying transversal $T$ to the physical qubits realizes the transversal logical $T$ (up to Pauli corrections).
\end{theorem}
\section{Asymptotically good CSS-T codes exist}\label{sec:CCSTconstruction}
The main goal of this section is to show the existence of asymptotically good CSS-T codes, thus answering the open question asked in \cite{RCNP20}.
We start by presenting a method to construct a CSS-T code from any given CSS code using an auxiliary map $\phi$. When $\phi$ is the identity map, we recover the construction and parameters presented in \cite{hu2021mitigating}. Moreover, when $\phi$ is a permutation map, we show that this construction yields the same parameters as in the identity case.
The CSS-T codes constructed using this method allow us to prove in \cref{th:goodcodes} and \cref{C:Asymp1} the existence of families of asymptotically good CSS-T codes. 
In particular, this construction preserves the LDPC property of the original CSS code, enabling us to show in \cref{T:Sparsity} and \cref{cor:asympldpc} the existence of asymptotically good quantum LDPC codes that are CSS-T.
\begin{definition}\label{def:phi2}
Let $C\subseteq\F_2^n$ be a linear code. Let $\phi:C\to\F_2^n$ be a map.
We define 
$$C^{N_\phi} \defeq \left\{(x,\phi(x)) \mid x\in C\right\}.$$
\end{definition}
In what follows, we focus on using a CSS code $(C_1,C_2)_\mathrm{CSS}$ and an appropriate map $\phi$ such that $\left(C_1^{N_\phi}, C_2^{N_\phi}\right)$ forms a CSS-T code. 
In the following proposition, we give necessary and sufficient conditions on the map $\phi$ so that $\left(C_1^{N_\phi}, C_2^{N_\phi}\right)$ forms a CSS-T code.
\begin{proposition}\label{prop:len-dim}
    Let $(C_1, C_2)_\mathrm{CSS}$ be a CSS code of length $n$. 
    Then $\left(C_1^{N_\phi}, C_2^{N_\phi}\right)$ forms a CSS-T code if and only if $\phi : C_1 \rightarrow \F_2^n$ is a linear map such that for each $x,y \in C_1$ and $z \in C_2$ we have $$\wt(x \star y \star z) + \wt(\phi (x) \star \phi (y) \star \phi (z))\equiv 0 \mod{2}.$$
\end{proposition}
\begin{proof} 
    By definition, $ C_2^{N_\phi} \subseteq  C_1^{N_\phi}$. Hence, $\left(C_1^{N_\phi}, C_2^{N_\phi}\right)$ is a CSS code if and only if  $ C_1^{N_\phi}$ and  $ C_2^{N_\phi}$ are linear codes, that is, 
    if and only if for any $x,y \in C_1$ we have $(x,\phi(x)) + (y,\phi(y)) = (x + y, \phi(x) + \phi(y)) = (x + y, \phi(x+y))$, where the second equality follows by the definition of $C_1^{N_\phi}$. Thus, $\phi$ must be linear on $C_1$. 
    Now, $\left(C_1^{N_\phi}, C_2^{N_\phi}\right)_\mathrm{CSS}$ is CSS-T if and only if $\left({C_1^{N_\phi}}\right)^{\star 2} \subseteq \left({C_2^{N_\phi}}\right)^\perp$ by \cref{th:csstchar}. So  $\left(C_1^{N_\phi}, C_2^{N_\phi}\right)_\mathrm{CSS}$ is CSS-T if and only if for each $x,y \in C_1$, $z\in C_2$ we have
    \begin{align*}
        ((x,\phi(x))\star (y,\phi(y)))\cdot (z,\phi(z)) &= (x\star y)\cdot z + (\phi(x)\star \phi(y))\cdot \phi(z) \\
        &= \sum_{i=1}^nx_iy_iz_i + \sum_{i=1}^n\phi(x)_i\phi(y)_i\phi(z)_i \\
        &= 0.
    \end{align*}
    The last equality holds if and only if $\wt(x \star y \star z) + \wt(\phi (x) \star \phi (y) \star \phi (z)) \equiv 0 \pmod{2}$. 
\end{proof}
\begin{proposition}\label{prop:par}
    Let $\phi$ be a map satisfying the conditions of \cref{prop:len-dim}. If $(C_1, C_2)_\mathrm{CSS}$ is an $[\![n,k]\!]$ CSS code, then  $\left(C_1^{N_\phi}, C_2^{N_\phi}\right)$ is a $[\![2n,k]\!]$ CSS-T code.
\end{proposition}
\begin{proof}
    This follows immediately by the properties of $\phi$ and the definition of $C^{N_\phi}$.
\end{proof}
Any map $\phi$ satisfying the conditions of \cref{prop:len-dim} allows constructing a CSS-T code from a CSS one. 
Let us point out that permutations of the coordinates, including the identity, satisfy those properties. Furthermore, using different permutations as the $\phi$ map will result in CSS codes with the same parameters, as shown in the following proposition.
\begin{proposition}\label{prop:permutations}
Let $C_2 \subseteq C_1 \subseteq \F_2^n$ be a pair of binary codes, and let $I$ and $\phi$ be the identity map and a permutation map on $\F_2^n$, respectively. Then, the CSS codes  
$\left(C_1^{N_I}, C_2^{N_I}\right)_\mathrm{CSS}$ and $\left(C_1^{N_\phi}, C_2^{N_\phi}\right)_\mathrm{CSS}$ have the same parameters. 
\end{proposition}
\begin{proof}
Clearly, the two CSS codes have the same length and dimension. So, we only need to deal with the minimum distance. Let $\Phi=(I,\phi)$ be the permutation of $\F_2^{2n}$ that leaves the first $n$ coordinates unchanged and applies $\phi$ to the second $n$ coordinates. Then $\Phi\left(C_1^{N_I}\right)=C_1^{N_\phi}$ and $\Phi\left(C_2^{N_I}\right)=C_2^{N_\phi}$. Therefore, $\Phi\left(C_1^{N_I} \setminus C_2^{N_I}\right)=C_1^{N_\phi} \setminus C_2^{N_\phi}$. Hence, both CSS codes have the same $X$-distance.  Since $\Phi$ is a permutation, we have 
$\Phi\left(\left({C_i^{N_I}}\right)^\bot\right)=\left(\Phi\left(C_i^{N_I}\right)\right)^\bot$ for $i \in \{1,2\}$. 
Thus
$\Phi\left(\left({C_i^{N_I}}\right)^\bot\right)=\left(\Phi\left(C_i^{N_I}\right)\right)^\bot=\left(C_i^{N_\phi}\right)^\bot$ and 
$\Phi\left({\left(C_2^{N_I}\right)}^\bot \setminus {\left(C_1^{N_I}\right)}^\bot\right)={\left(C_2^{N_\phi}\right)}^\bot \setminus {\left(C_1^{N_\phi}\right)}^\bot$
for $i \in \{1,2\}$. Hence, both CSS codes have the same $Z$-distance.  
\end{proof}
In the rest of the section, we fix $\phi$ to be the identity permutation $I$, and consider the code 
\begin{equation}\label{E:CN}
C^{N_I} \defeq \{(x,x) \mid x\in C\},
\end{equation}
which we will denote simply by $C^N$. This code coincides with the construction proposed in \cite[Section VII]{hu2021mitigating}, when taking $M=2$. Hence, relying on this previous work, we get the following result.
\begin{theorem}\label{T:CSST}
    For any given $[\![n,k,d]\!]$ binary CSS code $(C_1, C_2)_\mathrm{CSS}$, one can construct a $[\![2n, k,\geq d]\!]$ binary CSS-T code.
\end{theorem}
\begin{proof}
     Consider the CSS-T code $(C_1^N,C_2^N)$. By \cref{prop:par} we know that $\left(C_1^{N},C_2^{N}\right)_\mathrm{CSS-T}$ has length $2n$ and dimension $k$. 
 The bound on the minimum distance follows from \cite[Theorem 13]{hu2021mitigating}. 
\end{proof}
\begin{remark}\label{rmk:phi}
    Other choices of $\phi$ are possible and may lead to CSS-T codes with different parameters. As an example, consider the map $\phi$ defined by $\phi(x) = x + a$ for some $a \in C_1 \cap (C_1^{\star 2})^\perp$ and $x \in \mathcal{B}$ for some basis $\mathcal{B}$ of $C_1$. Finding maps $\phi$ giving CSS-T codes with better parameters than the choice $\phi=I$ or a permutation is an interesting open question.  We will return to it in the conclusion. 
\end{remark}
The next proposition shows that applying transversal $T$ to all the physical qubits of  CSS-T codes satisfying Theorem \ref{T:CSST} 
implies a logical identity operator. 
\begin{proposition}
Let $(C_1, C_2)_\mathrm{CSS}$ be a CSS-T pair. If $ C_1^{\star 2} \subseteq C_1^\perp$, then applying transversal $T$ to all the physical qubits implies a logical identity operator.    
In particular, all the CSS-T codes of Theorem \ref{T:CSST} satisfy this property.
\end{proposition}
\begin{proof}
The condition $ C_1^{\star 2} \subseteq C_1^\perp$ implies that $C_1 \subseteq C_1^\perp$, $C_1 \star C_2 \subseteq C_1^\perp$, $C_1 \star  C_1 \subseteq C_1^\perp$, and $C_2 \star C_2 \subseteq C_1^\perp$, which are conditions 1--4 of \cite[Theorem 12]{RCNP20b}, from which the result follows.
For the second part, let $\left(C_1^N, C_2^N\right)_\mathrm{CSS}$ be a CSS-T code of Theorem \ref{T:CSST}, where $C_i^N=(C_i,C_i)$, $i \in \{1,2\}$, for two length $n$ binary codes $C_1$ and $C_2$.
For each $x,y \in C_1$ and each $z\in C_1$ we have
    \begin{align*}
        ((x,x)\star (y,y))\cdot (z,z) &= (x\star y)\cdot z + (x\star y)\cdot z = 0.
    \end{align*}
    Thus $(C_1^N)^{\star 2} \subseteq (C_1^N)^\perp$ and this completes the proof.
\end{proof}
Table \ref{TL:1} provides a list of examples of binary CSS-T codes that we found after applying the construction of Theorem \ref{T:CSST} to binary linear cyclic codes. 
\begin{table}[ht]
\begin{center}
\begin{tabular}{|p{1.8 cm} p{1.8 cm} p{1.8 cm} p{1.8 cm}|}
\hline
\multicolumn{4}{|p{7.4cm}|}{\centering CSS-T 
 Parameters} \\
\hline
$[\![14, 3, 3]\!]$ & $[\![18, 2, 3]\!]$ 
& $[\![30, 1, 6]\!]$ & $[\![30, 2, 6]\!]$ \\
$[\![30, 8, 4]\!]$ & $[\![30, 10, 3]\!]$ &
$[\![42, 1, 6]\!]^\ast$
& $[\![42, 6, 6]\!]$ \\ 
$[\![42, 7, 5]\!]$ &
$[\![42, 12, 4]\!]$ & $[\![46, 1, 7]\!]$ 
&$[\![54, 1, 6]\!]^\ast$ \\
$[\![62, 1, 11]\!]$ &
$[\![62, 15, 6]\!]$ & $[\![66, 2, 10]\!]^\ast$ & \\
\hline
\end{tabular}
\end{center}
\caption{Small length binary CSS-T codes from cyclic codes. All the CSS-T codes are $Z$-degenerate. The codes shown by $\ast$ are also $X$-degenerate.} 
\label{TL:1}
\end{table}
\subsection{Asymptotically good families of CSS-T codes}
Recall that the rate and relative distance of a binary quantum code with parameters $[\![n,k,d ]\!]$ are defined as $R=\frac{k}{n}$ and $\delta=\frac{d}{n}$, respectively. 
In the literature, there exist several infinite families of quantum CSS codes with non-vanishing asymptotic rate and relative distance, commonly known as asymptotically good quantum codes, see for example \cite{ashikhmin2001asymptotically, chen2001asymptotically,panteleev2022asymptotically}. 
The existence of such families, combined with our Theorem \ref{T:CSST}, allows us to construct asymptotically good CSS-T codes.
\begin{theorem}\label{th:goodcodes}
From any given asymptotically good family of binary CSS codes, it is possible to derive an asymptotically good family of binary CSS-T codes.
\end{theorem}
\begin{proof}
Let $\{(C_1,C_2)_i\}$ be a family of asymptotically good CSS codes, where $i>0$ is an integer, with rate and relative distance $\frac{k_i}{n_i}$ and $\frac{d_i}{n_i}$, respectively, for each $i$. Let the 
asymptotic rate and relative distance be  $R=\displaystyle\lim_{i \to \infty} \frac{k_i}{n_i}>0$ and $\delta=\displaystyle\lim_{i \to \infty} \frac{d_i}{n_i}>0.$ 
Then applying the result of Theorem \ref{T:CSST} gives the infinite family
$\left\{\left({C_1}^{N},{C_2}^N\right)_i\right\}$ for each $i>0$ with 
asymptotic rate and relative distance $R'=\frac{R}{2}>0$ and $\delta'=\frac{\delta}{2}>0$, respectively.  
\end{proof}
\begin{corollary}\label{C:Asymp1}
Infinite families of quantum CSS-T codes with non-vanishing asymptotic rate and relative distance exist.  
\end{corollary}
\begin{proof}
Infinite asymptotically good families of binary CSS codes are presented in \eg \cite[Theorem 1]{ashikhmin2001asymptotically}, \cite[Theorem 1.2]{chen2001asymptotically}, and \cite[Theorem 2]{panteleev2022asymptotically}. The result then follows from Theorem \ref{th:goodcodes}.
\end{proof}
\subsection{Asymptotically good families of LDPC CSS-T codes}
Quantum codes with low-density parity check (LDPC) matrices are especially interesting in the context of quantum computing. 
Here we discuss how Theorem \ref{T:CSST} can yield asymptotically good quantum LDPC codes. 
The next theorem shows that applying the result of Theorem \ref{T:CSST} to a CSS code preserves its sparsity by sending its parity check matrix to another sparse parity check matrix. 
\begin{theorem}\label{T:Sparsity}
    Let $(C_1, C_2)_\mathrm{CSS}$ be a CSS code, and let
    $$H=\begin{bmatrix}
        H_X & 0\\
        0 & H_Z
    \end{bmatrix},$$
    where $H_X, H_Z$ generate $C_2$ and $C_1^\perp$, respectively. Then, $\left(C_1^N,C_2^N\right)_{\text{CSS-T}}$ has parity check matrix 
    $$H^N=\begin{bmatrix}
        H_X^N & 0\\
        0 & H_Z^N
    \end{bmatrix},$$
    where 
    \begin{center}   
    $H^N_X=\begin{bmatrix}
        H_X & H_X
    \end{bmatrix}$
    and
    $H^N_Z=\begin{bmatrix}
        H_Z & 0\\
        I & I
    \end{bmatrix}.$
        \end{center}
In particular, if $r_X$ and $r_Z$ are the maximum row weights of $H_X$ and $H_Z$, respectively, then  
the maximum row weights of $H_X^N$ and $H_Z^N$ are $2r_X$ and $\max\{r_Z,2\}$.
\end{theorem}
\begin{proof}
The definition of $C_2^N$ justifies that $H_X^N$ is a generator matrix for $C_2^N$. Moreover, one can easily check that the linear span of the rows of $H_Z^N$ is a subspace of $\left(C_1^N\right)^\bot$. Next, we show that the row space of $H_Z^N$ is $(C_1^N)^\bot$. Let $(x,y) \in \left(C_1^N\right)^\perp$ for some $x,y \in \F_2^n$. Let $e_i^N \defeq (e_i,e_i)$, where $e_i\in \F_2^n$ is the standard basis vector.
So there exists $\alpha_i \in \F_2$ for each $1\le i \le n$ such that 
$$(x,y) + \sum_{i=1}^n\alpha_ie_i^N =(x,y)+(y,y)=(x+y, 0) \in \left(C_1^N\right)^\bot.$$ 
This implies that $\forall a^N =(a,a) \in C_1^N,  (x+y, 0)\cdot a^N = 0$, hence  $\forall a\in C_1, (x + y) \cdot a = 0$, and thus  $x+y \in C_1^\bot$. 
Consequently, $(x+y, 0)$ is in the row space of $\begin{bmatrix}
        H_Z & 0
 \end{bmatrix}$. 
Since $(y,y)$ belongs to the row space of $\begin{bmatrix}
        I & I
 \end{bmatrix}$, $(x,y)$ is in the row space of $H_Z^N$. The structure of $H_X^N$ shows that its maximum row weight is $2r_X$. Moreover, the last $n$ rows of $H_Z^n$ have all weight $2$. The maximum weight of the remaining rows of $H_Z^N$ is $r_Z$. Thus, the maximum row weight of $H_Z^N$ is $\max\{r_Z,2\}$.
\end{proof}
There exist several families of good quantum LDPC codes, see \eg \cite{DinurAsymptoticallyGood,gooLDPChomological,LeverrierAsymptoticallyGood,hsiehGood,physicsGood1,physicsGood2}.
An immediate consequence of the above theorem is that applying the construction of Theorem \ref{T:CSST} to an LDPC CSS code preserves its sparsity. We use this property to entail the next result. 
\begin{corollary}\label{cor:asympldpc}
A family of asymptotically good quantum LDPC codes that are CSS-T exists.      
\end{corollary}
\begin{proof}
We apply the CSS-T construction of Theorem \ref{T:CSST} to each asymptotically good quantum LDPC family presented above. 
Theorem \ref{T:Sparsity} ensures that the resulting CSS-T codes are LDPC.  
\end{proof}
\section{Logical operators of different transversal gates}\label{sec:transversalgates}
In this section, we first develop the idea used in the previous chapter to produce asymptotically good CSS codes for any given $Z$ rotation. Then we discuss an application of them in mitigating coherent noise.
Secondly, we show that for any $(C_1,C_2)_\mathrm{CSS-T}$ with an even code $C_1$, the logical operator corresponding to the transversal $T$ has multiplicative order four. 
Therefore, the transversal $T$ gate never realizes transversal logical $T$ gate on such CSS codes.
Finally, we prove that the CSS-T codes from the construction of Theorem \ref{T:CSST} also support transversal $CCZ$ gate. In this case, the corresponding logical operator is the identity.
\subsection{Good CSS codes supporting transversal Z rotations}
As we already stated in \cref{C:Asymp1}, one can construct a family of asymptotically good CSS codes that supports transversal $T$ gate. 
Obviously, the same family supports other larger $Z$ rotations, namely transversal $Z$ or $S$ gate.
In this subsection, we discuss the construction of another asymptotically good CSS family that supports transversal application of other smaller $Z$ rotation gates. 
Next, we show that applying the construction of Theorem \ref{T:CSST} repeatedly preserves finer $Z$ rotations transversally.  
\begin{theorem}\label{thm:lconstruction}
Let $\ell > 3$ be an integer. Applying transversal $\varGamma^{(\ell)}$ gate to binary quantum CSS code $\left(C_{1}^{\ell}, C_{2}^{\ell}\right)$, obtained from applying Theorem \ref{T:CSST} $\ell$ times to $C_{2} \subseteq C_{1} \subseteq \F_2^n$, results in the logical identity operator.
\end{theorem}
\begin{proof}
Let $u \in \mathbb{F}_{2}^{k}$, where $k$ is the number of logical qubits in the quantum code, and $H$ be a binary matrix having a basis of $C_3^\ell$ in its rows, where $C_1^{\ell}= C_2^{\ell} \oplus C_3^{\ell}$. Note that the number of physical qubits is $2^{\ell} n$. 
Consider the logical state $\ket{u}_L=\sum_{v \in C_2^l}\ket{v+uH}$. 
Recall that the term $v + uH \in C_{1}^{\ell}$ for each selection of $u$ and $v$ and $\wt{(v + uH)}$ is the weight of corresponding vector in $C_{1}^{\ell}$. 
Moreover, since $C_{1}^{\ell}$ is obtained from applying the procedure of Theorem \ref{T:CSST} $\ell$ times, its codewords have weights divisible by $2^{\ell}$.
Thus
\begin{equation}\label{eq:eq1}
\begin{split}
\varGamma^{(\ell)^{\otimes{2^{\ell}n}}} \left( \sum_{v \in C_{2}^{\ell}} |v + uH\rangle \right) 
      &= \sum_{v \in C_{2}^{\ell}} \varGamma^{(\ell)^{\otimes{2^{\ell}n}}} |v + uH \rangle\\&=\sum_{v \in C_{2}^{\ell}} \left( \mathrm{e}^{\frac{2\pi i}{2^{\ell}}} \right)^{ \wt{(v + uH)}} |v + uH \rangle\\&=\ket{u}_L.
      \end{split}
\end{equation}
\end{proof}
The proof of the previous Theorem \ref{thm:lconstruction} can partially follow from \cite[Remark 13]{hu2022designing}. However, the latter does not provide any information about the structure of the logical operator. Note that one may get the same result after applying Theorem \ref{T:CSST} to $C_2 \subseteq C_1$ less than $\ell$ times. However, using $\ell$ repetitions keeps the proof more intuitive. An application of Theorem \ref{thm:lconstruction} is the following interesting result. The proof is similar to that of Corollary \ref{C:Asymp1}. So we skip it in here. 
\begin{corollary}
Let $\ell>0$ be a positive integer. Then, a family of asymptotically good CSS code exists that supports transversal $\varGamma^{(\ell)}$ gate.    
\end{corollary}
\subsubsection{Asymptotically good CSS codes for mitigating restricted coherent noise}
In quantum systems, coherent errors refer to systematic rotations around a fixed axis, which can accumulate over time, potentially leading to more severe damage than stochastic errors.
A natural and commonly studied model considers coherent noise acting transversally, where each qubit is subjected to an independent unitary operation \cite{hu2021mitigating}. 
This is the case, for example, when a uniform background magnetic field with the Hamiltonian $H = Z_1 + Z_2 + \cdots + Z_n,$
acts on the system, 
where $Z_i$ denotes the Pauli-$Z$ operator acting on the $i$-th qubit. 
Such interaction can induce a coherent error, corresponding to a $Z$-rotation by an angle $\frac{2\pi}{2^{\ell}}$, for some integer $\ell>0$, on each qubit. 
Although coherent noise can be addressed using active error correction techniques, a more resource-efficient approach is to passively mitigate its effects by encoding logical information into \emph{decoherence-free subspaces} (DFSs) \cite{alber2001stabilizing, kempe2001theory}. Particularly, in the language of stabilizer codes, the noise is required to preserve the code space and act as the logical identity.

The structure of CSS codes for which each arbitrary transversal $Z$-rotations act as the logical identity is discussed in \cite{hu2021mitigating}, and a construction for such codes is provided. These codes are commonly referred to as CSS codes that are \emph{oblivious to coherent noise}.

Let $\ell_{\max}\ge 3$ be an integer. In this paper, we consider
CSS codes where the code space
is preserved by transversal $\frac{2\pi}{2^{\ell}}$ $Z$-rotation, which acts as the identity operator on the code space, for each $\ell \le \ell_{\max}$. 
We refer to such codes as {\em oblivious to $\frac{2\pi}{2^{\ell_{\max}}}$-coherent noise}.
This approach is in line with the research direction presented in \cite{hu2021mitigating}, and in the following theorem we give a construction of asymptotically good quantum CSS codes that are oblivious to $\frac{2\pi}{2^{\ell_{\max}}}$-coherent noise, for any $\ell_{\max}$. 
\begin{theorem}
Let $\ell_{max} \ge 3$ be an integer. Then the binary quantum CSS code $\left(C_{1}^{\ell_{\max}}, C_{2}^{\ell_{\max}}\right)$, obtained from applying Theorem \ref{T:CSST} $\ell_{\max}$-times to $C_{2} \subseteq C_{1} \subseteq \F_2^n$ is oblivious to $\frac{2\pi}{2^{\ell_{\max}}}$-coherent noise.    
\end{theorem}
\begin{proof}
The proof of Theorem \ref{thm:lconstruction} shows that transversal $\frac{2\pi}{2^{\ell_{\max}}}$ $Z$-rotation implies the logical identity operator on the encoded space. 
Applying this transversal gate $2^{\ell_m-\ell}$ times, for each $\ell \le \ell_{\max}$, shows that again the corresponding logical operator is the identity, which is in correspondence to transversal $\frac{2\pi}{2^{\ell}}$ $Z$-rotation. Hence such CSS code is oblivious to $\frac{2\pi}{2^{\ell_{\max}}}$-coherent noise.  
$\frac{2\pi}{2^{\ell_{\max}}}$-coherent noise.    
\end{proof}
\subsection{Logical operator of CSS-T codes with even weights}
Recall that applying the $T$ gate transversally to all the physical qubits of a CSS-T code realizes a logical gate. 
In this subsection, we investigate the structure of this logical operator when both $C_2 \subseteq C_1$ are even and form a CSS-T code.  
We show that the multiplicative order of such a logical operator is always a divisor of four. 
In particular, such CSS-T codes never realize a transversal logical $T$ gate as the latter has the multiplicative order of eight.  
\begin{theorem}\label{T:ord}
Let $(C_1,C_2)$ be a CSS-T code with length $n$ such that $\wt(x)\equiv 0 \pmod 2$ for each $x\in C_1$. Then, the logical operator corresponding to $T^{\otimes n}$ has order a factor of four.        
\end{theorem}
\begin{proof}
Let $H$ be a binary matrix having a basis of $C_3$ in its rows, where $C_1= C_2 \oplus C_3$.
The CSS code of $(C_1,C_2)$ preserves the transversal $T$, implying that the following diagram commutes: 
\begin{equation}
\begin{tikzcd}[row sep=1.2cm, column sep=1.2cm]
\ket{u}_L \arrow[r,rightarrow, "\en"]  \arrow[d,"OP"]
& \displaystyle\sum_{v \in C_2}\ket{v+uH} \arrow[d, "T^{\otimes n}"] \\
\ket{u'}_L \arrow[r,rightarrow,"\en" ]&  \displaystyle\sum_{v \in C_2}\ket{v+u'H}=\displaystyle\sum_{v \in C_2}T^{\otimes n}\ket{v+uH}, 
\end{tikzcd}
\end{equation}
where $OP$ is the corresponding logical operator. 
Applying the operator $OP$ four times implies the next diagram
\begin{equation}\label{E:tra}
\begin{tikzcd}[row sep=1.2cm, column sep=1.2cm]
\ket{u}_L \arrow[r,rightarrow, "\en"]  \arrow[d,"OP^4"]
& \displaystyle\sum_{v \in C_2}\ket{v+uH} \arrow[d, "{T^4}^{\otimes n}"] \\
\ket{u''}_L \arrow[r,rightarrow,"\en" ]&  \displaystyle\sum_{v \in C_2}{T^4}^{\otimes n}\ket{v+uH}.
\end{tikzcd}
\end{equation}
Moreover, 
$${T^4}^{\otimes n}\ket{v+uH}= \mathrm{e}^{(\frac{8\pi i}{8}) \wt(v+uH)}\ket{v+uH}=\ket{v+uH}$$
as $\wt(v+uH)$ is even. 
So the bottom-left part of Equation \eqref{E:tra} should be the same as the top-left part (\ie $\ket{u}_L=\ket{u''}_L$), which implies that $OP^4$ is the logical identity.      
\end{proof}
Thus, all CSS-T codes obtained from the extended construction of classical codes and Reed--Muller (RM) (but not punctured RM) codes satisfy the conditions of Theorem \ref{T:ord}.  
\subsection{Application of transversal CCZ gate}
Let $u, v, w \in \mathbb{F}_2^{n}$. Recall that the transversal $CCZ$ gate is defined as follows:
\begin{equation*}
    CCZ^{\otimes n} |u\rangle |v\rangle |w\rangle \longmapsto (-1)^{\displaystyle\sum_{i = 1}^{n} (u_{i}v_{i}w_{i})}|u\rangle |v\rangle |w\rangle. 
\end{equation*}
In this subsection, we show that the quantum CSS code obtained from Theorem \ref{T:CSST} also supports the transversal CCZ gate. Recall that for each $C_2 \subseteq C_1$, we have set 
$C_i^N \defeq \{(z,z): z \in C_i\}$
for $i\in \{1,2\}$.
We also fix 
$H$ to be a binary matrix having a basis of $C_3^N$ in its rows, where $C_1^{N}= C_2^{N} \oplus C_3^{N}$.
\begin{theorem}
Let $C_2 \subseteq C_1 \subseteq \F_2^n$ be two binary codes. The quantum CSS-T code $\left(C_1^N,C_2^N\right)_\mathrm{CSS}$ has the following property: applying the transversal $CCZ$ gate to all $2n$ physical qubits realizes the logical identity gate.
\end{theorem}
\begin{proof}
Let $u, x, w \in \mathbb{F}_2^k$, where $k$ is the dimension of the mentioned CSS code. 
Applying the $CCZ$ gate to all the physical qubits sends the logical state $\ket{u}_L\ket{x}_L\ket{w}_L$ to the state
\begin{equation}\label{E:CCZ equi}
\begin{split}
    CCZ^{\otimes 2n} &\left( \sum_{v \in C_2^N} |v + uH \rangle \right) 
    \left( \sum_{v' \in C_2^N} |v' + xH \rangle \right) \\&
    \left( \sum_{v'' \in C_2^N} |v'' + wH \rangle \right).
\end{split}
\end{equation}
Expanding this gives:
\begin{equation*}
   CCZ^{\otimes 2n}\left(\sum_{v, v', v'' \in C_2^N} |v + uH \rangle |v' + xH \rangle |v'' + wH \rangle \right).
\end{equation*}
Now, applying $CCZ^{\otimes 2n}$ to each term in the summation gives:
\begin{equation*}
    \begin{split}
    &\sum_{v, v', v'' \in C_2^N} CCZ^{\otimes 2n} |v + uH \rangle |v' + xH \rangle |v'' + wH \rangle 
    \\&= \sum_{v, v', v'' \in C_2^N} CCZ^{\otimes 2n} |a \rangle |b \rangle |c \rangle \\&= \sum_{v, v', v'' \in C_2^N} (-1)^{\displaystyle\sum_{i=1}^{2n} a_i b_i c_i} |a\rangle |b\rangle |c\rangle\\&=\sum_{v, v', v'' \in C_2^N} (-1)^{\displaystyle 2\sum_{i=1}^{n} \alpha_i \beta_i \gamma_i} |a\rangle |b\rangle |c\rangle,
    \end{split}
\end{equation*}
where $a = v + uH=(\alpha,\alpha)$, $b = v' + xH=(\beta,\beta)$, and $c = v'' + wH=(\gamma,\gamma)$ are codewords of $C_1^N$ for some $\alpha,\beta,\gamma \in \F_2^n$. 
Thus, the result of (\ref{E:CCZ equi}) is  
\begin{equation*}
\left( \sum_{v \in C_2^N} |v + uH \rangle \right) 
    \left( \sum_{v' \in C_2^N} |v' + xH \rangle \right) 
    \left( \sum_{v'' \in C_2^N} |v'' + wH \rangle \right).
\end{equation*}
This implies that $CCZ^{\otimes2n}$ acts as the logical identity. 
\end{proof}
The above theorem can be extended to the case when there are more than two controls. However, we skip that case here. 
\section{New construction of triorthogonal codes}\label{sec:triorthogonal}
In this section, we discuss a method for transforming certain CSS-T codes into triorthogonal codes. 
Then, we apply this result to provide a general construction of triorthogonal codes using binary self-orthogonal codes (Theorem \ref{T:CSST to Triothogonal}), which is the main result of this section. Finally, we provide examples of triorthogonal codes constructed using our approach.

Since triorthogonal codes are a subclass of CSS-T codes, it is natural to search for a method of transforming CSS-T codes into them. This would allow using the magic state distillation protocol discussed in \cite{bravyi2012magic} for the transformed codes and find new applications of CSS-T codes. The following theorem gives an instance of such transformations.
\begin{theorem}\label{thm: cssttotriorth}
   Let $C_2 \subseteq C_1 \subseteq \mathbb{F}_{2}^{n}$ and $C_{1} = C_{2} \oplus \Span\{w_1, \ldots, w_m\} $ where each $w_i$ has an odd weight. Suppose that $\Span\{w_1, \ldots, w_r\}$ forms a triorthogonal subspace for some $r\leq m$. If $(C_1, C_2)$ is a CSS-T pair , then $C_2' = C_2 \subseteq C_1' = C_{2} \oplus \Span\{w_1, \ldots, w_r\}$  forms a triorthogonal CSS code.  
\end{theorem}
\begin{proof}
    Since $(C_1, C_2)$ is a CSS-T pair, we have $C_2^{\star 2} \subseteq C_2^{\bot}$, implying that  $C_2$ is a triorthogonal subspace. For  $j \leq r$, we have the following sequence of implications for each $c_2$ and $c_2' \in C_2$:
    \begin{equation*}\begin{split}
    &C_1^{\star 2} \subseteq C_2^\perp \implies (w_j\star c_2)\cdot c_2' = 0\pmod 2 \\&\implies \wt(w_j\star c_2 \star c_2') = 0 \pmod 2.\end{split}\end{equation*}
    Taking $c_2 = c_2'$, we get $\forall c_2\in C_2, \wt(w_j\star c_2) = 0 \pmod 2$. 
    On the other hand, the following holds as well for each $c_2\in C_2$ and $j,k\leq m$:
    $$ C_1^{\star 2} \subseteq C_2^\perp \implies (w_j\star w_k) \cdot c_2 = 0 \implies \wt(w_j\star w_k \star c_2) = 0 \pmod 2.$$
    Using the linearity property of these codes, one can extend the above results to any arbitrary vectors in $C_1$ and $C_2$, which proves the triorthogonality of the CSS code constructed from $C_2 \subseteq C_1$.
\end{proof}
\begin{corollary}\label{C:1-dim triorthogonal}
Let $C_2 \subseteq C_1 \subseteq \mathbb{F}_{2}^{n}$ form a CSS-T pair. Then the CSS code of $C_{1}' = C_{2} \oplus \Span\{w\} $, where $w\in C_1$ is a single vector of odd weight, and $C_2'=C_2$ is a triorthogonal code. 
\end{corollary}
\begin{proof}
    The result follows immediately since a single vector forms a triorthogonal subspace.
\end{proof}
Through the last part of this section, we generalize the doubling construction of triorthogonal codes to expand the spectrum of suitable codes for magic state distillation.
The doubling transformation was first discussed in \cite{betsumiya2012triply} and has been used to generate triply-even CSS codes, a class of codes closely related to triorthogonal codes but with some subtle differences. Recently, several triply-even CSS codes with optimal parameters have been constructed using this approach \cite{jain2024}. The main ingredients in the doubling construction are a doubly-even code and a triorthogonal code. In the following theorem, we relax the previous requirement by using a pair of self-orthogonal and triorthogonal codes.
This generalizes the earlier construction by enabling the production of triorthogonal codes that are not necessarily triply-even. We represent the all-one and all-zero vectors of length $n$ by $\bf{1_n}$ and $\bf{0_n}$, respectively.  
\begin{theorem}\label{T:CSST to Triothogonal}
Let $C_2$ be a binary self-orthogonal code of an odd length $n_1$ and $C_1=C_2\oplus\Span\{\bf{1_{n_1}}\}$ with the corresponding CSS parameters $[[n_1,1,d_1]]$. Let $C'_2 \subset C_1'$ be a triorthogonal CSS code with parameters $[[n_2,1,d_2]]$, where $n_2$ is also an odd integer. Then one can construct a triorthogonal code, not necessarily triply-even, with parameters
$[[2n_1 + n_2, 1, \min\{d_1, d_2 + 2\}]]$. 
\end{theorem}
\begin{proof}
Without loss of generality, we assume that $C_1'=C_2'\oplus \Span\{\bf{1_{n_2}}\}$ as otherwise the same proof works by replacing $\bf{1_{n_2}}$ with other relevant vector.
One can construct a pair of new binary codes, namely $C_2''\subset C_1''$, where $C_2''$ has the generator matrix \[G = \begin{bmatrix}
        C_2 & C_2&\bf{0_{n_2}}\\
        \bf{0_{n_1}}&\bf{0_{n_1}}&C_2'\\
        \bf{0_{n_1}}&\bf{1_{n_1}}&\bf{1_{n_2}}
    \end{bmatrix},
\] 
and $C_1''=C_2''\oplus \Span\{\bf{1_{2n_1+n_2}}\}$. The CSS code of $C_2'' \subseteq C_1''$ is a $[[2n_1+n_2,1, d]]$ code. Moreover, a similar proof as that of \cite[Lemma 3]{bravyi2015doubled} can be employed to show that $d=\min\{d_1, d_2 + 2\}$. 
Next, we show that $C_2'' \subseteq C_1''$ is a CSS-T pair. 
Since $C_2$ is a self-orthogonal code of an odd length, we have ${\bf{1_{n_1}}} \in C_2^\bot$ and $x \star y$ has an even weight for each $x,y \in C_2$. Then for each $u$ and $v \in C_2''$, the vector $u \star v$ has one of the forms 
\begin{enumerate}
    \item $u \star v=(x \star y,x \star y,{\bf{0_{n_2}}})$ for some $x,y \in C_2$,
    \item $u \star v=({\bf{0_{n_1}}},{\bf{0_{n_1}}},x' \star y')$ for some $x,y \in C_2'$,
    \item $u \star v=({\bf{0_{n_1}}},{\bf{1_{n_1}}},{\bf{1_{n_2}}})$.
    \item $u \star v=({\bf{0_{n_1}}},{x},{\bf{0_{n_2}}})$ for some $x \in C_2$,
\end{enumerate}
or a linear combination of them. Moreover, an arbitrary codeword of $C_2''$ is in the form of 
$$w=(a_1x_1,a_1x_1+a_3{\bf{1_{n_1}}},a_2x_1'+a_3{\bf{1_{n_2}}} ),$$
where $x_1\in C_2$, $x_1' \in C_2'$, and $a_i \in \F_2$ for $i=1,2,3$. 
An easy observation based on the forms (1)--(4) and the structure of $w$ shows that $(u \star v)\cdot w=0 \pmod 2$ for any $u,v,w \in C_2''$. 
Hence, we have $C_2'' \star C_2'' \subseteq C_2''^\bot$. 
Moreover, ${\bf{1_{2n_1+n_2}}} \star {\bf{1_{2n_1+n_2}}}={\bf{1_{2n_1+n_2}}} \in C_2''^\bot$, and for any $w \in C_2''$, we have $w \star {\bf{1_{2n_1+n_2}}}=w \in C_2''^{\bot}$. Hence we conclude that $C_1'' \star C_1'' \subseteq C_2''^{\bot}$ or equivalently $C_2'' \subseteq C_1''$ is a CSS-T pair.  
 Finally, such CSS-T pair satisfies the conditions of Corollary \ref{C:1-dim triorthogonal}, hence it is a triorthogonal code.    
\end{proof}
\cref{T:CSST to Triothogonal} allows for the construction of triorthogonal codes more easily, and enables a broader spectrum of classical codes to be used for this purpose. All the triorthogonal codes discussed in \cite{jain2024} can be obtained using the above theorem. Moreover, one can get codes beyond them. 
In the next section, we discuss how to extract suitable ingredients from the family of binary self-dual codes, following the strategy used in \cite{jain2024}, but focusing exclusively on singly even codes. 

Using this approach, one gets a broader list of codes for constructing triorthogonal codes as:
a binary doubly-even self-dual code of length $n$ satisfies $n\equiv 0 \pmod 8$ \cite[Page 31]{rains2002self}. However, a binary singly even self-dual code of length $n$ exists for each even integer $n$. 
Moreover, the number of singly even codes is much larger than the doubly-even codes. 
For instance, the number of singly-even (respectively, doubly-even) codes of lengths 24 and 32 are 46 (respectively, 9) and  3210 (respectively, 85), respectively. 
\subsection{Ingredient codes and examples}
In practice, obtaining a triorthogonal code appears to be more challenging than obtaining a binary self-orthogonal code. Therefore, to apply the result of Theorem \ref{T:CSST to Triothogonal}, it is more efficient to fix a triorthogonal code and search for quantum self-orthogonal codes. 
So, in what follows, we discuss a scenario for constructing quantum self-orthogonal codes.

Note that all the computations displayed in the following examples, including the actual minimum distances of our codes, were done using the Magma computer algebra system \cite{magma}.

Let $C$ be a binary self-dual code of length $n$ (which is thus an even integer) with minimum distance at least 2. Then $C=C^\bot$ and has parameters $[n,\frac{n}{2}]$. 
Shortening the code $C$ in any position gives $C_2$, which is a self-orthogonal code and it has parameters $[n-1,\frac{n}{2}-1]$. Then choosing $C_1=C_2 \oplus \Span\{\bf{1_{n-1}}\}$ leads to a pair $C_2 \subseteq C_1$ which is an $[[n-1,1]]$ CSS code. One can easily see that such a code satisfies the conditions of Theorem \ref{T:CSST to Triothogonal}.     

This approach replaces the requirement for a binary self-orthogonal code in Theorem \ref{T:CSST to Triothogonal} with a new ingredient, namely a binary self-dual code. Self-dual codes form an important class of classical codes, and several best-known codes and families of quantum codes have been constructed using them. For instance,  the parameters of several self-dual codes are calculated in \cite{gaborit2003experimental} (until length 130) and in \cite{dastbasteh2024new} (until length 241 mix of binary and quaternary). Moreover, there exist other databases of self-dual codes with good parameters, for instance \cite{Data1, Data2}, that allow easier integration of such codes to build triorthogonal codes. Next, we produce some examples of triorthogonal codes using the approach discussed above.   

Applying the result of Theorem \ref{T:CSST to Triothogonal} to the triorthogonal code $[[15,1,3]]$ (see \cite{bravyi2015doubled}), and the singly even self-dual codes of lengths $18$ and $20$ in \cite{Data2} gives triorthogonal codes with parameters 
\begin{enumerate}
    \item $[[49,1,3]]$ (degenerate to 2) and $[[49,1,5]]$ (degenerate to 4 and it was discussed in \cite{bravyi2015doubled}),
    \item $[[53,1,3]]$ (non-degenerate and degenerate to 2) and $[[53,1,5]]$ (degenerate to 4).
\end{enumerate}
In the following example, we consider a case where a singly even code yields a better triorthogonal code than those obtained from doubly-even codes with the same minimum distance.  
\begin{example}
Let $C$ be the linear binary cyclic code of length $89$ generated by the polynomial 
\[
\begin{split}
g(x)=&x^{45} + x^{44} + x^{42} + x^{38} + x^{36} + x^{35} + x^{33} + x^{32} +\\& x^{30} + x^{27} + x^{26} +
    x^{24} + x^{23} + x^{20} + x^{19} +x^{18} +  \\& x^{16} + x^{15} + x^{12} + x^8 + x^5 + x^4 +
    x^3 + 1.
\end{split}    
\]    
One can verify that $C$ is self-orthogonal and $C^\bot=C\oplus \Span \{{\bf{1_{89}}}\}$. Moreover, $C$ is singly even and cannot be used as an ingredient in the triorthogonal construction of \cite{jain2024}.
Our computation shows that the CSS code of $C \subseteq C^\bot$ has parameters $[[89,1,17]]$ (degenerate to 12). Then applying Theorem \ref{T:CSST to Triothogonal} to this code and the distance 15 triorthogonal code of \cite{jain2024}, namely $[[575, 1, 15]]$, produces a triorthogonal code with parameters $[[753,1,17]]$ which has better parameters than the minimum distance 17 triorthogonal code produced in \cite{jain2024}, namely $ [[777, 1, 17]]$.    
\end{example}
\section{Conclusion}\label{sec:conclusion}
In this paper, we proved the existence of asymptotically good CSS-T quantum (LDPC) codes, resolving an open problem proposed in \cite{RCNP20}. We also showed that the same holds for other finer $Z$ rotations. 
Our main technique was based on employing a construction of CSS-T codes integrated from concatenation and a mapping $\phi$. In the particular case of $\phi=I$, our technique overlapped that of \cite{hu2021mitigating}.
For the codes obtained from our approach, we also analyzed the logical operator of certain transversal gates. Particularly, we concluded that such logical operator is always the identity, and discussed the application of such codes for mitigating coherent noise.

We also introduced a novel doubling transformation that extends the doubling construction described in \cite{jain2024} for generating triorthogonal codes. Our approach accommodates self-orthogonal codes as the main component, instead of the doubly-even codes discussed in \cite{jain2024,bravyi2015doubled}. This generalization expands the class of codes suitable for constructing triorthogonal codes, thereby offering greater flexibility for magic state distillation.

In this paper, we chose the map $\phi$ to be the identity map as other permutation maps induce CSS-T codes with the same parameters. 
An interesting remaining question is to find other alternative maps $\phi$ satisfying the conditions of \cref{prop:len-dim} and leading to CSS-T codes with better parameters or new applications (see also \cref{rmk:phi}). 
As another future direction, we aim to systematically search for non-doubly-even self-orthogonal codes to serve as the main ingredient in our triorthogonal construction. This can be achieved through a study of certain classical codes and their parameters, with the goal of potentially reducing the overhead of magic state distillation using such codes.
\section*{Acknowledgment}
The order of the authors is purely alphabetical. All authors contributed to the paper to the same extent.

The authors thank Shubham P. Jain for an insightful private communication. They are also grateful to Narayanan Rengaswamy for his support and valuable advice that led to the final version of the paper.

\bibliographystyle{IEEEtran}
\bibliography{biblio_CSST}
\end{document}